# Fast Monte Carlo Dose Calculation in Proton Therapy

Jason Holmes, PhD[1*], Hongying Feng, PhD[2,1,3*], Lian Zhang, PhD[1], Michael Fix, PhD[4], Steve B. Jiang, PhD[5], Wei Liu, PhD[1]

[1]Department of Radiation Oncology, Mayo Clinic, Phoenix, AZ, USA, 85054

[2]College of Mechanical and Power Engineering, China Three Gorges University, Yichang, Hubei 443002, People's Republic of China

[3]Department of Radiation Oncology, Guangzhou Concord Cancer Center, Guangzhou, Guangdong, 510555, People's Republic of China

[4]Division of Medical Radiation Physics and Department of Radiation Oncology, Inselspital, Bern University Hospital, and University of Bern, Bern, Switzerland

[5]Medical Artificial Intelligence and Automation Laboratory and Department of Radiation Oncology, University of Texas Southwestern Medical Center, Dallas, TX, USA, 75390

[*]Co-first authors who contribute to this paper equally

Corresponding author: Wei Liu, PhD, Professor of Radiation Oncology, Department of Radiation Oncology, Mayo Clinic Arizona, 5777 E. Mayo Boulevard, Phoenix, AZ 85054; e-mail: Liu.Wei@mayo.edu.

**Acknowledgments**

This research was supported by the National Cancer Institute (NCI) Career Developmental Award K25CA168984, the Arizona Biomedical Research Commission Investigator Award, the Lawrence W. and Marilyn W. Matteson Fund for Cancer Research, and the Kemper Marley Foundation.

## Conflicts of Interest Notification

No conflicts of interest to report.

## Ethical considerations

This research was approved by the Mayo Clinic Arizona institutional review board (IRB, 13-005709). The informed consent was waived by IRB protocol. Only CT image and dose-volume data were used in this study. All patient-related health information was removed prior to the analysis and publication of the study.

## Data Availability Statement

The data are available from the corresponding author upon reasonable request.

**Abstract**

This article examines the critical role of fast Monte Carlo dose calculations in advancing proton therapy techniques, particularly in the context of increasing treatment customization and precision. As adaptive radiotherapy and other patient-specific approaches evolve, the need for accurate and precise dose calculations, essential for techniques like proton-based stereotactic radiosurgery, becomes more prominent. These calculations, however, are time-intensive, with the treatment planning/optimization process constrained by the achievable speed of dose computations. Thus, enhancing the speed of Monte Carlo methods is vital, as it not only facilitates the implementation of novel treatment modalities but also improves the optimality of treatment plans. Today, the state-of-the-art in Monte Carlo dose calculation speeds is $10^6$ - $10^7$ protons per second. This review highlights the latest advancements in fast Monte Carlo dose calculations that have led to such speeds, including emerging artificial intelligence-based techniques, and discusses their application in both current and emerging proton therapy strategies.

## Introduction

The standard delivery technique for proton therapy today is pencil beam scanning (PBS)[1-10]. PBS involves raster scanning proton beamlets across the tumor volume at discrete locations, spot by spot in the lateral dimension, layer by layer in the longitudinal dimension. The spot position within one energy layer is controlled by two orthogonal steering magnets while the proton range is controlled by changing the proton energy, either by the accelerator directly (synchrotron) or by incorporating energy degraders (cyclotron). The number of protons delivered to each spot position is determined by optimizing the spot weights such that dosimetric objectives are met by introducing dosimetric constraints appropriately together with machine-specific minimum monitor unit limits and by post-processing to adjust the optimized spot weights to be deliverable. These constraints aim to produce a dose distribution conformal to the tumor shape and uniform across the entire tumor volume while minimizing the dose to healthy tissues and in particular, organs at risk (OARs)[11-14].

The number of degrees of freedom in optimizing dose for cancer patients treated with PBS are vast. For this reason, as computational power has increased since the advent of PBS, the complexity of techniques used to deliver the optimized dose have increased in equal measure. Today, the standard treatment planning optimization technique used in PBS proton therapy is known as robust optimization[5,15-44]. In robust optimization, the dose is calculated for many potential real-world treatment perturbation scenarios[45,46] that include patient positioning errors, proton range errors, etc. For each perturbation scenario, $k$, the dose from each discrete spot, $j$, must be calculated once for each geometrical voxel, $i$, during the plan optimization. As an example, for robust optimization in a volume composed of 1,000,000 voxels irradiated by 1,000 spots with 10 perturbation scenarios considered, 10 matrices (per the number of perturbation scenarios), known as the dose influence matrix, $D_{ij}^{k}$ $(\mathrm{k} = 1,2,\ldots,10)$, are calculated for each spot, each matrix containing 1,000,000,000 elements. The spot weights are then optimized until the dosimetric constraints are met according to the robust optimization approach taken, broadly categorized in two ways, worst-case and stochastic. In the worst-case method, the worst-case perturbation scenario (maximum dose for OARs, minimum dose for target coverage, and maximum dose for target hot spot)[18] is optimized until the dosimetric constraints are met. In the stochastic method, the expected plan quality (weighted average among all scenarios) is optimized. With many optimization iterations, a treatment plan which is "robust" to the perturbation scenarios will emerge. Even as robust optimization is quite complex already, considering that in a time-resolved treatment (4D) that follows the respiratory motion of the patient, the number of dose calculations grows much further still.

An exciting and sophisticated development in the clinical workflow of proton therapy is the concept of adaptive radiation therapy[47-53]. In a typical course of treatment, the patient will need to make many visits to the hospital to be treated with a fraction of the total prescribed dose, known simply as a "fraction". Ideally, for each fraction, the patient staged for treatment would receive a CT scan and if necessary, the original treatment plan would be adjusted based on the new CT. Finally, the patient would be treated with the newly adapted plan. The plan adaption process, taking place after the CT and before treatment, should not put too much stress on the patient or radiation therapists and should not slow down the proton therapy clinic. Crucially, minimizing the duration of the adaptive process not only alleviates stress on both patients and radiation therapists but also significantly enhances the accuracy and effectiveness of the treatment plan. To allow for adaptive radiotherapy while incorporating other advanced optimization techniques such as 4D robust optimization, beam angle optimization, spot position optimization, linear-energy-transfer-based and relative biological effectiveness optimization[31,54-67], the efficiency of Monte Carlo dose calculation methods must be improved greatly.

While optimization and clinical techniques have grown in complexity, so too have dose calculation techniques. Because Monte Carlo-based dose calculations have historically been quite slow, treatment planning has typically been performed using analytical methods[68-71]. However, it is generally accepted that the most accurate dose calculation techniques utilize the Monte Carlo method. Additionally, for some emerging techniques in PBS proton therapy (PBSPT), such as dynamic collimation[72-75] and magnetic resonance (MR)-guided proton therapy[76], Monte Carlo methods are the only way to handle the relevant complicated particle transports.

In proton therapy, Monte Carlo dose calculation involves tracking individual protons and secondary particles step by step within the patient geometry. At each step, many interactions are possible with varying probabilities, based on physics. The Monte Carlo approach is to sample each independent interaction possibility, at each step, based on their respective probability, and apply the result. As more protons are simulated and the phase space of potential interactions for the entire treatment plan are sufficiently sampled, an accurate dose distribution will converge. The statistical uncertainty in dose for each voxel is therefore directly related to the number of protons which were simulated within the voxel.

General-purpose Monte Carlo simulations developed by the larger physics community have been in continual development for many years and even decades. The most well-known of these codes are Geant4[77], FLUKA[78], and MCNPX[79]. Many of the known radiation physics processes and their respective cross sections, built up over generations of experimentation across the world, have been compiled and inserted into these Monte Carlo codes. These codes allow for extreme precision and accuracy in calculations relating to radiation. In proton therapy, these codes are rightfully considered to be the gold standard for dose calculation. However, these codes are generally not viewed as being viable within the daily clinical workflow as they are much too slow, requiring hours, days, or even weeks to calculate the dose for a single proton therapy plan. For this reason, there has been much effort put into the development of fast Monte Carlo dose calculators specifically tailored for proton therapy[80-91] dose calculations. Typically, these specialized Monte Carlo codes are then validated against the gold standard general-purpose Monte Carlo codes. In this article, we will discuss the current status of Monte Carlo dose calculation methods in proton therapy followed by some recent developments which may enable the use of Monte Carlo dose calculation in the most demanding clinical workflows.

## Current Monte Carlo dose calculation methods in proton therapy

### A. Physics models of Monte Carlo dose calculation methods in proton therapy

The predominant types of interactions between protons and atoms or nuclei of media, with the proton energy in the therapeutic range (usually from 70 to 250 MeV[92]), include Coulombic interactions with atomic electrons, elastic and inelastic interactions with atomic electrons (ionization), Coulombic interactions with atomic nuclei, and inelastic nuclear interactions. Since protons undergo huge numbers of Coulombic interactions with either atomic electrons or atomic nuclei, it is impractical to simulate such Coulombic interactions one by one using Monte Carlo methods. Therefore, continuous models are used for Coulombic interactions. On the contrary, the occurrence of inelastic nuclear interactions, for example, is much less frequent and can therefore be considered as discrete events. Nonetheless, such discrete events are considerably more complex, requiring a plethora of physics models for the proton-nuclei interaction, which will often lead to the generation of secondary particles including $\delta$-electrons, positrons, neutrons, protons, deuterons, tritons, alphas, and gamma-rays, etc. Consequently, each discrete event requires independent addressment. Based on the aforementioned high-level diagram of the fundamental physics mechanism beneath the proton transport, a Class II Monte Carlo algorithm[93] is usually implemented to model the proton track in a step-by-step fashion, where associated physics processes are divided into condensed-history (continuous) models and point-like (discrete) models. The step length $d$ is determined using $d = \min(d_{vol}, d_{hard}, [d_{max}])$ (the bracket denotes optional variables). $d_{vol}$ is the translational distance between the current position of the proton and the next simulating volume interface, and $d_{hard}$ is the distance to the next discrete event. In voxelized simulating domains like Computed Tomography (CT) image, the simulating volume is the voxel, while in independent continuous simulating domains like range shifter, the simulating volume is the independent continuous simulating domain itself. $d_{hard}$ can be sampled based on the maximum interaction cross section ($\Sigma_{hard}^{max}$) by introducing the "fictitious" interaction cross section[80,82,83,94,95], sometimes referred to as Woodcock tracking. Alternatively, $d_{hard}$ can be determined by the real current total interaction cross section ($\Sigma_{hard}$) [81,84,96,97], which is dependent on the energy of incident proton and the media. Usually, a maximum step length $d_{max}$ can be introduced to constrain the step length[82,83,94].

According to ICRU Report 46, only a few elements are necessary to describe human tissues, e.g., Hydrogen, Carbon, Nitrogen, Oxygen, Phosphorus, and Calcium. Based on the element concentration, a truncation on the number of elements to be simulated can be predetermined, carefully balancing efficiency and accuracy. Human tissue related materials are constructed as a composition of the simulated elements, with those recommended by Schneider being the most commonly utilized[98]. These materials are then calibrated to the Hounsfield Unit (HU) in the CT image. The physical properties of each material that are necessary for dose calculation in proton therapy have been intensively studied by theoretical and experimental physicists, resulting in well tabulated datasets for direct queries and derivative interpolations.

### A.1 Continuous interactions

For the Coulombic interactions with the atomic electrons, protons are continuously slowing down, with production of $\delta$-electrons (ionizations) above the kinetic energy threshold $T_e^{min}$. In each step, protons lose an amount of energy, subject to energy straggling, described by the formula,

$$d = \int_{E_i - \Delta E}^{E_i} \frac{dE}{L(E)}, \tag{1}$$

where $E_i$ is the kinetic energy at the beginning of the step, $\Delta E$ is the energy loss. $L(E)$ denotes the restricted or unrestricted stopping power, depending on whether $\delta$-electrons are explicitly considered (> $T_e^{min}$) or not. In the restricted scenario, the effective stopping power of $\delta$-electrons production process[80,99], i.e., the first moment of the energy differential macroscopic cross section of $\delta$-electrons production process, is subtracted from the unrestricted stopping power. The unrestricted and restricted stopping power were well described by the Bethe equation with corrections[97,99]. PSTAR[100], Geant4, and International Committee for Radiological Units (ICRU) Report 49 provide tabulated stopping powers of common materials for convenient query and calculations. Alternatively, the stopping power of a material can be expressed by the water stopping power with a correction i.e., the stopping power ratio of the material to water[80]. The $\delta$-electrons (no matter explicitly addressed or not) can be considered to deposit energies locally for most scenarios, since the maximum energy $T_e^{max}$ transferred from the most energetic therapeutic protons to $\delta$-electrons corresponds to an electron range around 2 mm in water, which is similar to voxel sizes commonly used for dose calculation (2~3 mm). $T_e^{max}$ was given by the formula,

$$T_e^{max} = \frac{2m_e\beta^2\gamma^2}{1 + 2\gamma m_e/m_p + \left(m_e/m_p\right)^2}, \tag{2}$$

where $m_e$ and $m_p$ are the electron and proton rest masses, $\beta$ is the ratio of proton velocity to the velocity of light ($c$), $\gamma$ is the relativistic parameter given by $\gamma = E_p/m_p$ with $E_p$ denoting the total proton energy. However, when it comes to air-inflated tissues like lungs, the such $\delta$-electrons can travel a larger distance, which requires an implicit consideration of the electron transport to further increase the calculation accuracy[80,99].

The production of the $\delta$-electrons in the continuously slowing down model of protons (i.e., electrons with subthreshold energies) leads to energy fluctuations of the primary protons (i.e., energy straggling). Two different models of fluctuations are applied depending on the thickness of the absorber material, which is determined by a parameter $\kappa = \Delta E/T_e^{max}$ with $\Delta E$ denoting the mean continuous energy loss that can be calculated by Eq. (1)[97] and approximated by only considering the leading term in the Bethe-Bloch model for $L(E)$[101]. For thick absorbers ($\kappa > 10$), which is the case for most applications in proton therapy, energy straggling behaves according to a Gaussian distribution with standard deviation $\delta E$[80] added to $\Delta E$ according to Bohr's theory[102]. For thin absorbers, energy straggling is well described by the Landau-Vavilov model[103], which can be fitted to a log-normal distribution for efficient sampling[104].

Due to the large rest mass ratio between proton and electron, therapeutic protons influenced by Coulombic interactions with the atomic electrons nearly travel a straight line. In contrast, when passing close to the atomic nucleus, protons will be elastically scattered or defected by the repulsive force from the positive charge of the nucleus. In proton therapy, most objects of interest are thick enough to produce a huge number of scattering events (i.e., MCS, multiple Coulomb scattering), whose net influence on the passing proton is a scattering angle ($\Delta\theta$) with negligible energy loss. The overall scattering angle of the MCS is well described by the Moliere's theory[105]. A zero mean, $\theta_0$ width Gaussian distribution is an excellent approximation for central 98% of the full Moliere distribution[84,99], which corresponds to the small-angle events. The calculation formula for the width $\theta_0$ was initially proposed by Rossi and Greisen[106], and later modified by Highland[107]. However, large-angle events can result in a long tail in the Moliere distribution, which cannot be sufficiently reproduced by the Gaussian or even the Gaussian mixture (multiple Gaussian, usually double) distribution. Therefore, a Rutherford-like distribution is added to the central Gaussian distribution to account for the wide tails[84,96,108].

### A.2 Discrete interactions

With respect to the continuous interactions, the discrete interactions (mainly elastic and nonelastic nuclear interactions for therapeutic protons) are characterized by short-range hard interactions that cause a more profound impact on the proton transport. ICRU 63 provided explicit definitions for elastic, nonelastic, and inelastic, but an equivalent use of the terms nonelastic and inelastic is commonly seen in the literature, meaning the kinetic energy is not conserved in the process. For instance, the target nucleus may undergo break-up, it may be excited into a higher quantum state, or a particle transfer reaction may occur. The occurrence of a discrete event is determined by $d_{hard}$ setup method. If the "fictitious" strategy is used, $d_{hard}$ is calculated by the formula $d_{hard} = -\ln(\eta)/\Sigma_{hard}^{max}$, where $\eta$ is a random number sample from uniform distribution from zero to unity[82,94,95]. If $d_{hard}$ is the decisive constraint for the step length $d$ ($d = d_{hard}$), then another number $\xi$ is sampled from a uniform distribution from zero and unity and determined whether $\xi$ is less than $\Sigma_{hard}/\Sigma_{hard}^{max}$. If yes, a discrete event occurs. If the $d_{hard}$ is determined by the real total cross section, $d_{hard}$ is sampled by the formula $d_{hard} = -\ln(\eta)/\Sigma_{hard}$ at the very beginning or just after a discrete event[84,97]. If $d_{hard}$ is larger than the actual step length $d$, the current step length $d$ will be subtracted from $d_{hard}$ until $d_{hard}$ becomes the decisive constraint for $d$, triggering a discrete event. For cases where $d_{hard}$ is not the decisive constraint for $d$, or $d_{hard}$ is the decisive constraint for $d$ yet a discrete event is rejected, the proton is simply transported with continuous interactions (see A. 1) within step length $d$. Once a discrete event happens, the specific type is determined according to the corresponding contribution to the total cross section and the proton undergoes the determined type of discrete event following the transport with continuous interactions (see A. 1) within step length $d$.

To reach the nucleus and trigger nuclear interactions, protons need to have adequate kinetic energy to overcome the repulsive Coulomb potential of the nucleus. The elastic nuclear interaction is described in a two-body scheme and the kinetics is usually solved in the center of mass (CM) system. For the *proton-proton elastic interactions*, the partial-wave analysis database Scattering Analysis Interactive Dial-in (SAID)[109] tabulated the microscopic cross sections that can be used to calculate the macroscopic cross sections. Meanwhile, Fippel and Soukup[80] generated a formula to calculate the macroscopic cross section based on the SAID database using analytical fitting. Alternatively, total and differential proton-proton elastic cross sections can be calculated by parameterization of Cugnon et al[81,110]. The angular distribution of protons after the collision is almost isotropic in the CM system, therefore it can first be sampled from a uniform distribution in the CM system and then transformed to the laboratory system. For the *proton-nucleus elastic interactions,* ICRU 63 and Evaluated Nuclear Data File (ENDF)[111] provided total and double differential (proton energy and scattering angle) cross sections for commonly used elements in proton therapy. Tripathi et al[112] also proposed a method for calculating proton-nucleus elastic cross sections. Scattering angles in the CM system can alternatively be sampled according to the parameterization of the elastic differential cross sections proposed by Ranft[81,96,113]. The scattered protons (both protons in the proton-proton elastic interactions) can be continuously simulated as the primary protons, while the heavier recoils deposit the transferred energy from the incident protons locally due to the large mass ratio between the recoils and protons, leading to negligible transport range compared to the voxel size[86]. It is worth noting that, large angle scattering (Rutherford tails) can either be considered in the cross section used here, or in the MCS previously mentioned. Additional attention is required to avoid the double counting of the large angel scattering.

During *proton-nucleus nonelastic interactions*, nuclear interactions can lead to a variety of reaction products, including neutrons, protons, deuterons, tritons, alphas, gamma-rays, and heavier fragments. To take full consideration of all these effects requires complicated algorithms like the Geant4 Binary Cascade and pre-compound models[97]. General-use Monte Carlo codes can separate the dose or energy deposition by particle species, allowing for a deeper understanding of the radiation effects. However, in the context of proton therapy, specific simplifications can be made while retaining a high degree of accuracy in terms of therapeutic dose calculation. Neutrons are usually neglected due to the trivial contribution to the local dose distribution[80,111,114]. Heavy fragments, having such a short range, do not require tracking and can be adequately simulated by depositing energy locally. The angle and emission energy of protons, deuterons, tritons, and alphas can be directly sampled from the double-differential cross sections given in ICRU 63. The secondary protons can be treated the same as primary protons. In the simplest approximation, secondary deuterons, tritons, and alphas deposit energy locally[80,82]. In less simplified models, such secondaries (not necessarily all of them) are simulated as if they were protons with energy and mass corrections to conserve energy and the proton-equivalent range[82,115]. More explicit simulations of such secondaries can also be carried out in the same manner as protons using the ENDF database. Prompt gamma-rays, which may be of fundamental importance for proton range verification[116] and imaging[117,118] techniques, are typically neglected when only proton dose is desired, however their inclusion in Monte Carlo dose calculations may become important if prompt gamma-ray detection techniques are implemented for clinical use. The emission of prompt gamma-rays can be sampled from the double-differential cross sections given in ICRU 63 with an isotropic angular distribution[115]. More details on the transport of prompt gamma-rays using the Monte Carlo method can be found in the PENELOPE[95] code reference.

## B. Monte Carlo-based Dose Calculation and Robust Optimization of a PBS proton therapy plan

In the dose calculation and robust optimization (essentially the calculation of the influence matrices) of a PBSPT plan, protons are simulated starting from the exit of the treatment gantry nozzle before any beam modification devices. Within the whole simulating domain, two different coordinates are used, i.e., the beam eye view (BEV) coordinate corresponding to the configuration of each treatment field (field angle, field isocenter, and lateral scanning position of each spot) and the associated devices (range shifters and apertures), and the CT coordinate corresponding to the patient setup. In both the BEV and CT coordinates, the governing physics models for proton transport are the same, described in the previous sections. When protons finish the propagation in the BEV coordinate (range shifters, apertures, or air gaps) and reaches the surface of the patient body, a coordinate transformation is carried out based on the configuration of each treatment field and the CT coordinate will be used in the patient body. A common difference between the proton simulation outside and inside the patient body is that the simulation inside the patient body is voxelized with dose and linear energy transfer (LET) scoring[119], while the simulation outside the patient body may treat beam modifying devices as coherent 3D objects where the dose and LET are not necessarily scored.

For dose calculation and robust optimization, the dose resolution (i.e., voxel size) is usually not adopted from the CT image resolution, but instead usually set to be 2 to 3 mm, sometimes 1 mm in stereotactic body radiation therapy (SBRT)[120]. As a result, to correctly score the dose and LET within a dose voxel, either the CT can be resampled to be identical to the dose resolution via HU number interpolation[121] or the scored dose in a dose voxel can be weighted by the volume or mass of nearby CT voxels (i.e., CT voxels that have volume overlapping with the dose voxel).

**B.1 Dose Calculation of a PBSPT plan**

In the dose calculation of a PBSPT plan, the proton energy, lateral scanning position, and the Monitor Unit (MU) of each beamlet are pre-defined through institution-specific machine properties and treatment planning. The dose calculation is carried out beamlet by beamlet, therefore dose calculation of one beamlet is taken as an example for the following detailed implementation. At the beginning, the initial phase space of protons is randomly sampled according to the spatial distribution of the proton beamlet measured/modelled during the beam commissioning process. Then the trajectory of each proton is simulated through the models described in previous sections, with dose or energy deposition scored within each voxel. The corresponding secondary protons to be simulated are treated in the same workflow as the primary proton, with their initial phase space sampled from the nonelastic nuclear interaction model where they are generated. For other secondaries except proton, certain approximations or additional simulations are needed according to physics models described in Section A. In a Monte Carlo simulation of the dose distribution of one primary proton, the simulation should be repeated enough times to achieve reasonably low statistical uncertainty (usually around 1% in the target voxels for a plan with all beamlets considered). Then the averaged dose distribution of one primary proton can be calculated, which is then multiplied by the number of protons from the beamlet determined by the weight of the beamlet (MU number), which is converted to number-of-protons based on the MU-proton number conversion curve. After finishing the dose calculation for each beamlet, a summation over the dose distributions of all beamlets is done to obtain the final dose distribution of a PBSPT plan.

**B.2 Robust Optimization of a PBSPT plan**

In the clinical practice of robust optimization of a PBSPT plan, the fundamental procedure is to calculate the influence matrices[122], $D_{i,j}$, i.e., the contribution of *j-th* spot with unit intensity at *i-th* voxel in the region of interests (ROIs), which can be considered as spot-by-spot dose distributions in proton therapy. Though the basic dose calculation method is the same as the method in the dose calculation of a PBSPT plan, a few differences need to be emphasized. First, in the robust optimization, constraints to shape the dose distribution are structure-based. Therefore, the voxels to be considered are the Boolean summation of the contained voxels of each selected structure (i.e., the region of interests), where constraints are placed upon, usually resulting in a much smaller number of dose voxels than the total dose voxels in the case of dose calculation of a PBSPT plan. Second, the beamlets (energy and position) to be used are not pre-determined but can be selected via raytracing to the target volume. Third, the number of protons within each selected beamlet (i.e., MU) is not pre-determined, but is set to be unit intensity across all selected beamlets during the calculation of influence matrices, which are then used to optimize the MU for each beamlet, according to the constraints, yielding the optimized dose distribution in terms of target coverage and OARs sparing.

As for the PBSPT plan robustness[123-125], the patient setup uncertainties and the proton range uncertainties[45,46] are usually considered in the robust optimization. For patient setup uncertainties, the isocenter of each treatment field is usually shifted a distance (usually 3-5 mm), up and down in three cardinal directions. For proton range uncertainties, the stopping powers (or stopping power ratios) of all the materials are scaled up and down usually by 3%. Then the patient setup and proton range uncertainties are combined to construct a space of perturbation scenarios (including the nominal one without uncertainties considered). For each robust scenario, a corresponding influence matrix can be calculated. Worst-case robust optimization[15,18] is one of the most widely used methods where for each voxel, the worst-case dose value (maximum dose for OARs, minimum dose for target coverage, and maximum dose for target hot spot) is select to evaluate the plan quality and guide the robust optimization process. While we have presented common practices above, note that more advanced/comprehensive considerations could be done per the clinical case and the institution's capability, such as random setup uncertainties, respiratory motion uncertainties, and delivery specific uncertainties[126] (aperture positioning or aperture shape), etc.

**Recent Developments and Applications**

**A. Graphic Processing Unit Acceleration**

Graphic Processing Unit (GPU) with the Compute Unified Device Architecture (CUDA) framework is a commonly used tool to accelerate Monte Carlo simulations in PBSPT[81,83,87,88,127,128], while Open Computing Language (OpenCL) framework is also used[84]. In CUDA, every 32 threads are grouped into one warp, in which the same instructions are executed for each thread simultaneously. However, if control flow branches (if… else…) exist, the runtime of the threads may diverge, significantly reducing the parallel efficiency[87].

In the Monte Carlo simulation, there are generally two particle tracking strategies, event-based[128,129] and history-based. In the event-based technique, particle history is split into basic components (such as continuous interactions, ionizations, elastic nuclear interactions and nonelastic nuclear interactions), which are first accumulated and then processed by different corresponding kernels. This technique is friendly for GPU acceleration and is inherently immune to the thread divergence problem. Although event-based techniques avoid the divergence problem, they suffer from global memory latency[129]. In contrast, the history-based technique is more widely adopted, in which the particle history is continuously tracked until the termination condition is satisfied. Since proton histories differ, with secondary particle generation being the biggest branch, the thread divergence problem is inevitable.

Several techniques have been developed to address the divergence problem in GPU-accelerated proton dose engines implemented in a history-based fashion. In gPMC[83], protons were simulated in batches with a size of *M*, with a special stack created to store secondary protons. In each batch, *M* protons were simulated while the daughter protons were stored in the stack. All other secondary particles were not tracked and were locally deposited for simplification. When the stack contains *M* or more daughter protons, the stack would pop up *M* protons to GPU to be simulated in the following batch. The gMC method[81] introduced two loops of particle simulation; primary protons are simulated in the first loop, and secondary protons generated are stored and processed only after all primary protons have been simulated. This process repeats until all secondary protons are simulated. Lastly, MOQUI[88] adopted a similar strategy to gMC and gPMC for managing thread divergence by queuing secondary particles for later simulation, but it innovated by utilizing a hash-table to efficiently manage the limited GPU memory, allowing for the scoring of quantities that would otherwise require extensive memory.

## B. Track-Repeating Algorithm

Track-repeating is a concept to sample particle tracks from a pre-calculated database of particle histories instead of on-the-fly calculation in Monte Carlo simulation of particles not only for protons[89], but also for photons, electrons, and carbons[130,131]. To generate the tracks, one could use protons of multiple energies (including the highest) or the protons of the highest energy only to balance the memory consumption and simulation time. For each step in the track, a complete set of particle's information without degeneracy can be recorded, such as phase space, energy at the start of the step and energy deposit within the step[131] or the transport length, angles relative to the previous step, energy loss, and energy deposit[85]. The tracks can be generated using only water phantom, complemented by modification of the particle's information during the repeating of one selected track[85], or using different materials so as to select the track corresponding to the material at the location of the particle[131]. Since only the tracks of protons of a few (or the maximum) energies are pre-generated, whereas protons of a wide range of energies (compared to the tracked protons) are used in a PBSPT plan, the starting step corresponding to protons of a certain energy used in a PBSPT plan needs to be searched out to truncate the corresponding tracks. For this, the "in-track search" method can be used[131], potentially done during pre-processing. The tracks of secondaries (usually only secondary protons) are also recorded, and secondaries can be treated the same as primary protons. Such a technique is very GPU-friendly, since by assigning the same proton history within a CUDA block only with different starting positions in the normal direction of the beam, each GPU thread essentially performs the same operations all the time.

## C. Virtual Particle Monte Carlo

Virtual Particle (VP) is a novel concept proposed as a counterpart of realistic particle, the particle conventionally considered in a Monte Carlo simulation of the proton therapy, i.e., the primary protons and the corresponding secondaries generated during the tracking history of the primary protons[87]. VP is a statistical concept that equivalently converted the histories of realistic particles (i.e., primary and secondary protons with further simplifications) to the histories of VPs, in terms of particle transport and energy deposit (thus dose and linear transfer energy (LET) calculation[132,133]).

Each VP corresponded to one proton (either primary or secondary). In a conventional Monte Carlo simulation, primary protons are initialized at the beginning while secondary protons are generated during the tracking histories of primary protons according to certain possibility distribution functions (PDFs) determined by the characteristics of the penetrating proton and penetrated medium. However, in a VP Monte Carlo (VPMC) simulation, all VPs are initialized at the beginning, and the complexity and randomness of secondary proton generation are eliminated. Therefore, the governing models and controlling logic are simplified to be identical for each VP in a VPMC simulation, which is optimal for CUDA-parallelization.

Pre-calculated physics parameters (the deposited energy, energy straggling, the deflection angle, weight, and the ionization probability) databases (i.e., PDFs), generated based on the simulation records of realistic particles using a fast Monte Carlo dose engine in phantoms with different materials, are also used to further increase the calculation efficiency in VPMC simulations, by using database querying instead of on-the-fly calculation taking advantage of CUDA's powerful capability of texture.

### D. Beamlet-free dose optimization

As previously discussed, current methods for dose optimization typically include the calculation of the dose influence matrix for a number of perturbation scenarios, followed by optimizing the intensity of each beamlet. These influence matrices can be extremely large, requiring significant memory. Additionally, the matrix multiplications required during gradient descent of the objective function can be costly, reducing the speed of the optimization. Although the optimization process itself is typically separated from the Monte Carlo dose calculations when utilizing the dose influence matrix, it is also possible to optimize the spot weights during the dose calculation[134]. A recently proposed method, known as "beamlet-free optimization"[135], eliminates the dose influence matrix by combining the dose calculation and optimization process, thereby requiring far less memory (95% reduction) and reducing the overall time from plan creation to the final optimized dose calculation by up to 75% for complex clinical cases.

The beamlet-free dose optimization method utilizes the same cost function as conventional methods, typically including maximum and minimum dose constraints of ROIs and the targets. However, rather than optimizing the intensities of individual beamlets, the beamlet-free algorithm optimizes the dose directly by sampling the cost function during the simulation. This is achieved by simulating a small number of protons at random spot locations and estimating the gradient of the cost function (the difference of the cost function before and after). Given the gradient estimate at the spot position, the spot weight (number of protons to be delivered) is adjusted towards minimizing the gradient. This process is similar to the stochastic gradient descent method. Once optimization is complete, after sampling the cost function sufficiently throughout the dose volume, the result is the final optimized dose. Unlike dose influence matrix-based optimization methods, no additional final dose calculation or aggregation is required. Although memory demand is not necessarily a concern for modern computation systems, this method may become more important to allow for increasingly complex optimization methods in the future.

### E. AI-based MC dose calculation and denoising

AI-based image processing can achieve near real-time, ultra-high-quality outputs in many applications[136,137,138]. Benefiting from AI's rapid progress in image processing, AI-based Monte Carlo dose generation has gradually become a popular research topic in recent years[139]. The ultimate goal is to achieve super-fast or even real-time high-precision Monte Carlo dose generation with the aid of AI. AI-based Monte Carlo dose generation can be categorized into two classes: 1) AI-based Monte Carlo dose calculation: This involves determining the precise dose based on a specific set of machine parameters and patient anatomy. 2) AI-based Monte Carlo dose denoising: This method utilizes AI to convert noisy low-statistic Monte Carlo doses to high-statistic Monte Carlo doses.

AI-based Monte Carlo dose calculation utilizes machine parameters and patient imaging data, typically CT images, as inputs. This approach is more closely aligned with the functionality of traditional Monte Carlo dose engines. Neishabouri et al. demonstrated a long short-term memory (LSTM) network which achieved high accuracy in proton Monte Carlo dose calculations with up to 98.57% γ-index passing rate, and offered a substantial reduction in calculation times ranging from 6 to 23 ms[140]. Zhang et al. introduced a novel deep learning-based DiscoGAN framework for Monte Carlo dose calculation in proton therapy, achieving consistent performance across various treatment sites and beam energies[141]. Pastor-Serrano et al. presented a deep learning algorithm DoTA that calculates proton therapy doses with high accuracy, achieving a 99.37% gamma pass rate compared to Monte Carlo simulations and delivering results in 5 ms[142]. Wu et al. developed a deep learning model that converts the low-precision doses calculated by a pencil beam algorithm to high-precision doses calculated by Monte Carlo methods for proton therapy across multiple disease sites[143].

Another approach in AI-based Monte Carlo dose generation is AI-based dose denoising. This involves using noisy low-statistic doses as inputs, which are then converted into high-precision doses calculated by Monte Carlo methods through AI -based denoising. Bai et al. developed a real-time, deep learning-based dose denoiser plugin to covert the noisy low-statistic Monte Carlo dose to high-statistic Monte Carlo dose, enabling the entire calculation time to be completed in 0.15 s[144]. Further studies have reported the application of AI-based Monte Carlo dose denoising in MRI-guided radiotherapy[145], proton therapy[146,147], carbon-ion radiotherapy[148], and also CT imaging dose[149].

Dose generation is a key component in radiation therapy planning, especially in the context of adaptive radiotherapy, involving iterative plan design/finetuning, re-planning, and rapid plan quality assurance[150,151]. AI-based Monte Carlo dose calculation and denoising techniques have demonstrated remarkable capabilities in providing ultrafast computation speeds and high-precision dose outputs. These advancements enable the rapid generation of high-precision Monte Carlo doses for future online adaptive radiotherapy. Recently, dose prediction is an emerging research area in AI-based dose generation[152-154], but it’s crucial to note that AI-based dose prediction is distinct from AI-based dose calculation and denoising. Dose calculation and denoising refer to determining the precise dose based on a specific set of machine parameters and patient anatomy. In contrast, dose prediction involves determining an optimal dose distribution for a given patient's anatomy. However, the pursuit of AI-based optimal dose prediction, particularly those achieving Monte Carlo-level accuracy, remains a noteworthy direction for research[155,156].

### F. Apertures in Monte Carlo dose calculations

Apertures are difficult to simulate for two main reasons, their extreme density as compared to normal tissues or bones, and their upstream position, which amplifies poorly approximated aperture interactions downstream due to geometric scaling. Because protons may interact with the aperture and still reach the patient, they cannot be well approximated in analytical approaches. Monte Carlo-based dose calculation methods are currently the preferred methods for simulating apertures in hadron therapy[157]. It is for this reason, as well as due to an increasing interest in apertures used in pencil beam scanning hadron therapy, that fast Monte Carlo dose calculators should support the inclusion of apertures[72,158-165]. It should be noted that the use of apertures will result in neutron production, however neutrons are not simulated in proton therapy-specific dose calculators. Rather, where there is concern due to neutron production, it is common practice to investigate neutron production in one of the general-use Monte Carlo codes for a subset of patients.

The conventional approach to simulating apertures in proton therapy is to voxelize the aperture, i.e., converting the aperture opening cross section into planar voxels. In principle, the aperture can be accurately simulated using a voxelization approach so long as any features in the aperture opening are much larger than the size of the aperture voxels. For small aperture openings, this method can potentially become problematic and inefficient. The voxelization of apertures is intrinsically not well-aligned with how aperture-openings are defined since they are typically defined by an ordered list of points forming a closed polygon in the planes normal to the beam direction. Another method of simulating apertures is to determine whether particles are within the aperture or not based on the crossing number algorithm[157,158]. This method simulates the aperture in the same geometric manner in which it is defined, avoiding voxelization and using the minimal amount of information to define the aperture geometry, making it a fast and efficient method as well as easily modelling small apertures precisely.

### G. Reducing run times of general-use Monte Carlo physics codes

There is a long history of medical physicists implementing general-use Monte Carlo codes such as MCNPX, FLUKA, and Geant4, for radiation therapy applications[166-171] that continues to this day. While these codes are typically viewed as gold standards in terms of dose calculation accuracy, their clinical use has typically been limited to commissioning, verification, or evaluation of clinical software due to long run times. Because these codes track the history of individual particles, allowing for the generation of secondary particles, they are not well suited for acceleration via GPU processing (see Section A). On the other hand, Monte Carlo dose calculations are intrinsically well suited to parallelization using "embarrassingly parallel" methods – distributing the simulation of primary particles across separate nodes (threads, processors, physical nodes, virtual nodes), tracking them to their end, then joining the results from each distributed workload – since no communication is necessary between each processor. Perhaps due to the proliferation of fast, proton therapy-specific Monte Carlo dose calculators, or due to general-use Monte Carlo methods not being well suited for GPUs as-is, there has been relatively little effort put into reducing run times of general-use Monte Carlo codes for the purpose of radiation therapy dose calculation. However, we will highlight some research that has been conducted to this end.

General-use Monte Carlo codes are not merely dose calculators – rather, they are physics simulators, able to account for many radiation-related effects and secondary particles that are mostly neglected by modern, fast Monte Carlo dose calculators (as discussed before), and may therefore take on more important roles as the precision of radiation therapy continues to advance. A few platforms have been envisioned for reducing run times of general-use Monte Carlo codes, notably – MPEXS, a Geant4-based GPU dose engine[172], however perhaps the most attractive method today is cloud computing. Cloud computing can provide on-demand access to 10s or 1000s of virtual nodes for computation, establishing a pay-per-use cost model as opposed to purchasing and maintaining on-site computer hardware. The main appeal for cloud computing in the context of general-use Monte Carlo codes is that their underlying code does not require modification. The approach for dose calculation in the cloud is essentially the "embarrassingly parallel" approach, splitting up the total number of protons amongst many virtual machines, each running the general-use Monte Carlo code, then accumulating the results once each node is done. Using this approach, Keyes et. al. (2010)[173] were able to simulate about $1x10^4$ protons per second in a water phantom with FLUKA, including the time to distribute the simulation parameters (patient geometry and plan information) to each node as well as simulation initialization time. Green et. al. (2015)[174] were able to simulate $2.7x10^4$ protons per second for a realistic plan using Geant4, end to end. Finally, Wang et. al.[175] used cloud computing with FLUKA to study prompt gamma spectroscopy in the context of proton therapy.

## Discussion and outlook

Table 1: Monte Carlo-based proton calculation speeds for select studies published since 2015.

| Reference | Year | Method | Notes | Protons/s |
|---|---|---|---|---|
| Clinical validation of a GPU-based Monte Carlo dose engine of a commercial treatment planning system for pencil beam scanning proton therapy[176] | 2021 | GPU | Voxels: 1-3 mm side-lengths, 100s of patients | 8.4x10^6 |
| Virtual particle Monte Carlo: A new concept to avoid simulating secondary particles in proton therapy dose calculation[87] | 2022 | GPU | Voxels: 2.5 mm side-lengths, 13 patients | 2.9x10^7* *virtual-particles/s |
| A fast GPU-based Monte Carlo simulation of proton transport with detailed modeling of nonelastic interactions[81] | 2015 | GPU | Voxels: 1.0 mm side-lengths, 3 H&N patients, fast version | 4.8x10^5 |
| Fast multipurpose Monte Carlo simulation for proton therapy using multi- and many-core CPU architectures[177] | 2016 | CPU multi-threaded | Voxels: 1.0 mm side-lengths, heterogeneous phantom | 4.4x10^5 |
| Commissioning of GPU–Accelerated Monte Carlo Code Fred for Clinical Applications in Proton Therapy[178] | 2021 | GPU | Voxels: 1.5 mm side-lengths, 90 H&N/brain patients | 2.9x10^5 |
| Development and Benchmarking of a Monte Carlo Dose Engine for Proton Radiation Therapy[179] | 2021 | CPU multi-threaded | Voxels: 2.0 mm side-lengths, brain patient | 1.9x10^5 |
| MOQUI: an open-source GPU-based Monte Carlo code for proton dose calculation with efficient data structure[88] | 2022 | GPU | 1 H&N, 1 liver, and 1 prostate patient | 4.3x10^5 |
| Fast Monte Carlo proton treatment plan validation in the Google Cloud[174] | 2015 | Cloud | Voxels: 1x1x2 $mm^3$, H&N phantom | 5.0x10^4 |

This review article has presented a detailed overview of the current state of Monte Carlo methods used in proton therapy dose calculation, robust optimization with Monte Carlo dose engines, and recent advances in further increasing speeds. Table 1 provides the reported proton calculation rates of various Monte Carlo-based dose calculators. The reported speeds depend on many factors including the number of voxels, number of GPU or CPU cores used, the desired accuracy, etc., and should therefore not be considered a fair comparison, nor exhaustive. The purpose of Table 1 is to provide context on the overall state-of-the-art regarding proton calculation rates. Taking this context into account, the current state-of-the-art for Monte Carlo dose calculation speeds is $10^6$-$10^7$ protons per second. The primary motivations for increasing speeds today are robust optimization and adaptive radiotherapy, however many treatment techniques on the horizon, including FLASH, grid therapy, 4D planning, linear-energy-transfer/relative biological effectiveness optimization, as well as the general trend towards increasing fraction doses, all require increasingly accurate and precise dose calculations in addition to faster speeds[55,180-186]. The many recent developments discussed in this article are causing the field to quickly approach a point where Monte Carlo methods may completely overtake analytical approaches. The rise of artificial intelligence is introducing new opportunities in speeding up Monte Carlo dose calculations as well.

In terms of new techniques that have been developed for fast Monte Carlo dose calculations, accuracy and speed are often inversely correlated. Clinically, these competing concepts, accuracy and speed, must be delicately balanced depending on the application. This is less true when new hardware-centric methods are developed for increasing dose calculation speed. In the case of GPU-based methods, accuracy is sometimes sacrificed to make the calculations more suited for GPU processing, however not always. “Embarrassingly parallel” methods can increase speeds without sacrificing accuracy, albeit typically at a higher monetary cost. In general, hardware-based methods increase the cost of dose calculation in order to increase speeds. While AI methods can greatly speed up dose calculations, they are currently not generalizable, unlike conventional Monte Carlo approaches. A dose calculator using AI will only be valid for scenarios that were well represented in the training data. Therefore, the tradeoff with utilizing AI based methods in the context of dose calculation is an increase in speed at the cost of generalizability. It should be noted, however, that AI methods will likely increase in generalizability with time. With regards to increasing the speed of Monte Carlo dose calculations, there are many factors and tradeoffs that must be carefully considered.

An important question going forward: At what point may we consider dose calculations to be fast enough? Our answer, in short, is that Monte Carlo dose calculations will not be considered fast enough for the foreseeable future. This is primarily due to the high degree of freedom in dose optimization. For example, "robust" in robust optimization is typically referring to robustness with respect to patient translations and proton beam range uncertainties. However, we could also make plans robust to dose engines, HU or material mapping, simulation techniques, etc. As of today, these ideas may be considered too time consuming. However, in the limit where the time to process Monte Carlo-based dose calculations approach zero, we would surely consider many additional robustness scenarios. Furthermore, there are many parameters that are not typically considered for optimization today, due to the long calculation times that would be required. For example, beam angles, number of beams, spot positions, or even the shape of aperture openings, could be better optimized in a more comprehensive fashion. In general, the high degree of freedom in dose optimization means that there will always be opportunity for increasingly complex and high-quality optimization techniques. Additionally, considering recent developments in adaptive radiotherapy, increasing speeds for Monte Carlo-based dose calculators will continue to be an important aspect of proton therapy for the foreseeable future.

## Reference

1. Mutter RW, Choi JI, Jimenez RB, et al. Proton Therapy for Breast Cancer: A Consensus Statement From the Particle Therapy Cooperative Group Breast Cancer Subcommittee [published online ahead of print 2021/05/29]. *Int J Radiat Oncol Biol Phys.* 2021;111(2):337-359.
2. Jimenez RB, Hickey S, DePauw N, et al. Phase II Study of Proton Beam Radiation Therapy for Patients With Breast Cancer Requiring Regional Nodal Irradiation [published online ahead of print 2019/08/27]. *J Clin Oncol.* 2019;37(30):2778-2785.
3. Schild SE, Rule WG, Ashman JB, et al. Proton beam therapy for locally advanced lung cancer: A review. *World journal of clinical oncology.* 2014;5(4):568-575.
4. Liu W, Inventor. System and Method For Robust Intensity-modulated Proton Therapy Planning. 09/02/2014, 2014.
5. Zaghian M, Cao W, Liu W, et al. Comparison of linear and nonlinear programming approaches for "worst case dose" and "minmax" robust optimization of intensity-modulated proton therapy dose distributions [published online ahead of print 2017/03/17]. *J Appl Clin Med Phys.* 2017;18(2):15-25.
6. Frank SJ, Cox JD, Gillin M, et al. Multifield Optimization Intensity Modulated Proton Therapy for Head and Neck Tumors: A Translation to Practice. *International Journal of Radiation Oncology Biology Physics.* 2014;89(4):846-853.
7. Yu NY, DeWees TA, Liu C, et al. Early Outcomes of Patients With Locally Advanced Non-small Cell Lung Cancer Treated With Intensity-Modulated Proton Therapy Versus Intensity-Modulated Radiation Therapy: The Mayo Clinic Experience. *Advances in Radiation Oncology.* doi: 10.1016/j.adro.2019.08.001.
8. Yu NY, DeWees TA, Voss MM, et al. Cardiopulmonary Toxicity Following Intensity-Modulated Proton Therapy (IMPT) Versus Intensity-Modulated Radiation Therapy (IMRT) for Stage III Non-Small Cell Lung Cancer. *Clinical Lung Cancer.* 2022;23(8):e526-e535.
9. Shan J, Yang Y, Schild SE, et al. Intensity-modulated proton therapy (IMPT) interplay effect evaluation of asymmetric breathing with simultaneous uncertainty considerations in patients with non-small cell lung cancer [published online ahead of print 2020/09/24]. *Med Phys.* 2020;47(11):5428-5440.
10. Zhang X, Liu W, Li Y, et al. Parameterization of multiple Bragg curves for scanning proton beams using simultaneous fitting of multiple curves [published online ahead of print 2011/11/17]. *Phys Med Biol.* 2011;56(24):7725-7735.
11. Bhangoo RS, DeWees TA, Yu NY, et al. Acute Toxicities and Short-Term Patient Outcomes After Intensity-Modulated Proton Beam Radiation Therapy or Intensity-Modulated Photon Radiation Therapy for Esophageal Carcinoma: A Mayo Clinic Experience [published online ahead of print 2020/10/22]. *Adv Radiat Oncol.* 2020;5(5):871-879.
12. Bhangoo RS, Mullikin TC, Ashman JB, et al. Intensity Modulated Proton Therapy for Hepatocellular Carcinoma: Initial Clinical Experience [published online ahead of print 2021/08/20]. *Adv Radiat Oncol.* 2021;6(4):100675.
13. Yu NY, DeWees TA, Liu C, et al. Early Outcomes of Patients With Locally Advanced Non-small Cell Lung Cancer Treated With Intensity-Modulated Proton Therapy Versus Intensity-Modulated Radiation Therapy: The Mayo Clinic Experience [published online ahead of print 2020/06/13]. *Adv Radiat Oncol.* 2020;5(3):450-458.
14. Yu NY, DeWees TA, Voss MM, et al. Cardiopulmonary Toxicity Following Intensity-Modulated Proton Therapy (IMPT) vs. Intensity-Modulated Radiation Therapy (IMRT) for Stage III Non-Small Cell Lung Cancer. *Clinical Lung Cancer.* 2022. doi: https://doi.org/10.1016/j.cllc.2022.07.017.

15. Pflugfelder D, Wilkens JJ, Oelfke U. Worst case optimization: a method to account for uncertainties in the optimization of intensity modulated proton therapy. *Physics in Medicine and Biology.* 2008;53(6):1689-1700.
16. Unkelbach J, Bortfeld T, Martin BC, Soukup M. Reducing the sensitivity of IMPT treatment plans to setup errors and range uncertainties via probabilistic treatment planning. *Medical Physics.* 2009;36(1):149-163.
17. Fredriksson A, Forsgren A, Hardemark B. Minimax optimization for handling range and setup uncertainties in proton therapy. *Medical Physics.* 2011;38(3):1672-1684.
18. Liu W, Zhang X, Li Y, Mohan R. Robust optimization in intensity-modulated proton therapy. *Med Phys.* 2012;39:1079-1091.
19. Liu W, Li Y, Li X, Cao W, Zhang X. Influence of robust optimization in intensity-modulated proton therapy with different dose delivery techniques. *Med Phys.* 2012;39.
20. Chen W, Unkelbach J, Trofimov A, et al. Including robustness in multi-criteria optimization for intensity-modulated proton therapy. *Physics in Medicine and Biology.* 2012;57(3):591-608.
21. Liu W, Liao Z, Schild SE, et al. Impact of respiratory motion on worst-case scenario optimized intensity modulated proton therapy for lung cancers. *Practical Radiation Oncology.* 2015;5(2):e77-86.
22. Unkelbach J, Alber M, Bangert M, et al. Robust radiotherapy planning. *Physics in Medicine & Biology.* 2018;63(22):22TR02.
23. An Y, Liang JM, Schild SE, Bues M, Liu W. Robust treatment planning with conditional value at risk chance constraints in intensity- modulated proton therapy. *Medical Physics.* 2017;44(1):28-36.
24. An Y, Shan J, Patel SH, et al. Robust intensity-modulated proton therapy to reduce high linear energy transfer in organs at risk. *Medical Physics.* 2017;44(12):6138-6147.
25. Liu C, Patel SH, Shan J, et al. Robust Optimization for Intensity-Modulated Proton Therapy to Redistribute High Linear Energy Transfer (LET) from Nearby Critical Organs to Tumors in Head and Neck Cancer [published online ahead of print 2020/01/29]. *Int J Radiat Oncol Biol Phys.* 2020. doi: 10.1016/j.ijrobp.2020.01.013.
26. Liu CB, Schild SE, Chang JY, et al. Impact of Spot Size and Spacing on the Quality of Robustly Optimized Intensity Modulated Proton Therapy Plans for Lung Cancer. *International Journal of Radiation Oncology Biology Physics.* 2018;101(2):479-489.
27. Liu W, ed *Robustness quantification and robust optimization in intensity-modulated proton therapy.* Springer; 2015. Rath A, Sahoo N, eds. Particle Radiotherapy: Emerging Technology for Treatment of Cancer.
28. Liu W, Schild SE, Chang JY, et al. Exploratory Study of 4D versus 3D Robust Optimization in Intensity Modulated Proton Therapy for Lung Cancer. *International Journal of Radiation Oncology Biology Physics.* 2016;95(1):523-533.
29. Shan J, An Y, Bues M, Schild SE, Liu W. Robust optimization in IMPT using quadratic objective functions to account for the minimum MU constraint. *Medical Physics.* 2018;45(1):460-469.
30. Feng H, Shan J, Ashman JB, et al. Technical Note: 4D robust optimization in small spot intensity-modulated proton therapy (IMPT) for distal esophageal carcinoma [published online ahead of print 2021/06/01]. *Med Phys.* 2021. doi: 10.1002/mp.15003.
31. Feng H, Shan J, Anderson JD, et al. Per-voxel constraints to minimize hot spots in linear energy transfer (LET)-guided robust optimization for base of skull head and neck cancer patients in IMPT. *Med Phys.* 2021.
32. Zaghian M, Lim G, Liu W, Mohan R. An Automatic Approach for Satisfying Dose-Volume Constraints in Linear Fluence Map Optimization for IMPT [published online ahead of print 2014/12/17]. *J Cancer Ther.* 2014;5(2):198-207.

33. Shan J, Sio TT, Liu C, Schild SE, Bues M, Liu W. A novel and individualized robust optimization method using normalized dose interval volume constraints (NDIVC) for intensity-modulated proton radiotherapy [published online ahead of print 2018/11/06]. *Med Phys.* 2018. doi: 10.1002/mp.13276.
34. Liu C, Bhangoo RS, Sio TT, et al. Dosimetric comparison of distal esophageal carcinoma plans for patients treated with small-spot intensity-modulated proton versus volumetric-modulated arc therapies [published online ahead of print 2019/05/22]. *J Appl Clin Med Phys.* 2019;20(7):15-27.
35. Liu C, Sio TT, Deng W, et al. Small-spot intensity-modulated proton therapy and volumetric-modulated arc therapies for patients with locally advanced non-small-cell lung cancer: A dosimetric comparative study [published online ahead of print 2018/10/18]. *J Appl Clin Med Phys.* 2018;19(6):140-148.
36. Liu C, Yu NY, Shan J, et al. Technical Note: Treatment planning system (TPS) approximations matter - comparing intensity-modulated proton therapy (IMPT) plan quality and robustness between a commercial and an in-house developed TPS for nonsmall cell lung cancer (NSCLC) [published online ahead of print 2019/09/10]. *Med Phys.* 2019;46(11):4755-4762.
37. Liu W, Li Y, Li X, Cao W, Zhang X. Influence of robust optimization in intensity-modulated proton therapy with different dose delivery techniques. *Medical Physics.* 2012;39(6Part1):3089-3101.
38. Liu W, Mohan R, Park P, et al. Dosimetric benefits of robust treatment planning for intensity modulated proton therapy for base-of-skull cancers. *Practical Radiation Oncology.* 2014;4:384-391.
39. Liu W, Patel SH, Harrington DP, et al. Exploratory study of the association of volumetric modulated arc therapy (VMAT) plan robustness with local failure in head and neck cancer [published online ahead of print 2017/05/16]. *J Appl Clin Med Phys.* 2017;18(4):76-83.
40. Liu W, Patel SH, Shen JJ, et al. Robustness quantification methods comparison in volumetric modulated arc therapy to treat head and neck cancer. *Practical Radiation Oncology.* 2016;6(6):E269-E275.
41. Zhang P, Fan N, Shan J, Schild SE, Bues M, Liu W. Mixed integer programming with dose-volume constraints in intensity-modulated proton therapy [published online ahead of print 2017/07/07]. *J Appl Clin Med Phys.* 2017;18(5):29-35.
42. Feng H, Patel SH, Wong WW, et al. GPU-accelerated Monte Carlo-based online adaptive proton therapy: A feasibility study. *Medical Physics.* 2022;49(6):3550-3563.
43. Feng H, Sio TT, Rule WG, et al. Beam angle comparison for distal esophageal carcinoma patients treated with intensity-modulated proton therapy [published online ahead of print 2020/10/16]. *J Appl Clin Med Phys.* 2020;21(11):141-152.
44. Li H, Zhang X, Park P, et al. Robust optimization in intensity-modulated proton therapy to account for anatomy changes in lung cancer patients. *Radiotherapy and Oncology.* 2015;114(3):367-372.
45. Lomax AJ. Intensity modulated proton therapy and its sensitivity to treatment uncertainties 1: the potential effects of calculational uncertainties. *Physics in Medicine and Biology.* 2008;53(4):1027-1042.
46. Lomax AJ. Intensity modulated proton therapy and its sensitivity to treatment uncertainties 2: the potential effects of inter-fraction and inter-field motions. *Physics in Medicine and Biology.* 2008;53(4):1043-1056.
47. Yan D, Vicini F, Wong J, Martinez A. Adaptive radiation therapy. *Physics in Medicine and Biology.* 1997;42(1):123-132.
48. Yan D, Wong J, Vicini F, et al. Adaptive modification of treatment planning to minimize the deleterious effects of treatment setup errors. *International Journal of Radiation Oncology Biology Physics.* 1997;38(1):197-206.

49. Jagt T, Breedveld S, van de Water S, Heijmen B, Hoogeman M. Near real-time automated dose restoration in IMPT to compensate for daily tissue density variations in prostate cancer. *Physics in Medicine & Biology.* 2017;62(11):4254.
50. Bernatowicz K, Geets X, Barragan A, Janssens G, Souris K, Sterpin E. Feasibility of online IMPT adaptation using fast, automatic and robust dose restoration. *Physics in Medicine & Biology.* 2018;63(8):085018.
51. Nenoff L, Matter M, Charmillot M, et al. Experimental validation of daily adaptive proton therapy. *Physics in Medicine & Biology.* 2021;66(20):205010.
52. Zhang M, Westerly DC, Mackie TR. Introducing an on-line adaptive procedure for prostate image guided intensity modulate proton therapy. *Physics in Medicine and Biology.* 2011;56(15):4947-4965.
53. Botas P, Kim J, Winey B, Paganetti H. Online adaption approaches for intensity modulated proton therapy for head and neck patients based on cone beam CTs and Monte Carlo simulations. *Physics in Medicine & Biology.* 2019;64(1):015004.
54. An Y, Shan J, Patel SH, et al. Robust intensity-modulated proton therapy to reduce high linear energy transfer in organs at risk [published online ahead of print 2017/10/05]. *Med Phys.* 2017;44(12):6138-6147.
55. Deng W, Yang Y, Liu C, et al. A Critical Review of LET-Based Intensity-Modulated Proton Therapy Plan Evaluation and Optimization for Head and Neck Cancer Management [published online ahead of print 2021/07/22]. *Int J Part Ther.* 2021;8(1):36-49.
56. Liu C, Patel SH, Shan J, et al. Robust Optimization for Intensity Modulated Proton Therapy to Redistribute High Linear Energy Transfer from Nearby Critical Organs to Tumors in Head and Neck Cancer [published online ahead of print 2020/01/29]. *International journal of radiation oncology, biology, physics.* 2020;107(1):181-193.
57. Paganetti H. Mechanisms and Review of Clinical Evidence of Variations in Relative Biological Effectiveness in Proton Therapy. *International Journal of Radiation Oncology*Biology*Physics.* 2021. doi: https://doi.org/10.1016/j.ijrobp.2021.08.015.
58. Yang Y, Muller OM, Shiraishi S, et al. Empirical relative biological effectiveness (RBE) for mandible osteoradionecrosis (ORN) in head and neck cancer patients treated with pencil-beam-scanning proton therapy (PBSPT): a retrospective, case-matched cohort study. *Frontiers in Oncology.* 2022;12.
59. Yang Y, Patel SH, Bridhikitti J, et al. Exploratory study of seed spots analysis to characterize dose and linear-energy-transfer effect in adverse event initialization of pencil-beam-scanning proton therapy. *Medical Physics.* 2022;49(9):6237-6252.
60. Yang Y, Rwigema JM, Vargas C, et al. Technical note: Investigation of dose and LET(d) effect to rectum and bladder by using non-straight laterals in prostate cancer receiving proton therapy [published online ahead of print 20221018]. *Med Phys.* 2022;49(12):7428-7437.
61. Yang Y, Vargas CE, Bhangoo RS, et al. Exploratory Investigation of Dose-Linear Energy Transfer (LET) Volume Histogram (DLVH) for Adverse Events Study in Intensity Modulated Proton Therapy (IMPT) [published online ahead of print 2021/02/24]. *International journal of radiation oncology, biology, physics.* 2021;110(4):1189-1199.
62. Cao W, Khabazian A, Yepes PP, et al. Linear energy transfer incorporated intensity modulated proton therapy optimization [published online ahead of print 2017/11/14]. *Phys Med Biol.* 2017;63(1):015013.
63. Giantsoudi D, Grassberger C, Craft D, Niemierko A, Trofimov A, Paganetti H. Linear Energy Transfer-Guided Optimization in Intensity Modulated Proton Therapy: Feasibility Study and Clinical Potential. *International Journal of Radiation Oncology*Biology*Physics.* 2013;87(1):216-222.

64. Inaniwa T, Kanematsu N, Noda K, Kamada T. Treatment planning of intensity modulated composite particle therapy with dose and linear energy transfer optimization [published online ahead of print 2017/03/24]. *Phys Med Biol.* 2017;62(12):5180-5197.
65. Traneus E, Oden J. Introducing Proton Track-End Objectives in Intensity Modulated Proton Therapy Optimization to Reduce Linear Energy Transfer and Relative Biological Effectiveness in Critical Structures [published online ahead of print 2018/11/06]. *International journal of radiation oncology, biology, physics.* 2019;103(3):747-757.
66. Bai X, Lim G, Grosshans D, Mohan R, Cao W. Robust optimization to reduce the impact of biological effect variation from physical uncertainties in intensity-modulated proton therapy [published online ahead of print 2018/12/14]. *Phys Med Biol.* 2019;64(2):025004.
67. Unkelbach J, Botas P, Giantsoudi D, Gorissen BL, Paganetti H. Reoptimization of Intensity Modulated Proton Therapy Plans Based on Linear Energy Transfer [published online ahead of print 2016/11/22]. *International journal of radiation oncology, biology, physics.* 2016;96(5):1097-1106.
68. Hong L, Goitein M, Bucciolini M, et al. A pencil beam algorithm for proton dose calculations. *Physics in Medicine and Biology.* 1996;41(8):1305-1330.
69. Schaffner B, Pedroni E, Lomax A. Dose calculation models for proton treatment planning using a dynamic beam delivery system: an attempt to include density heterogeneity effects in the analytical dose calculation. *Physics in Medicine and Biology.* 1999;44(1):27-41.
70. Younkin JE, Morales DH, Shen J, et al. Clinical Validation of a Ray-Casting Analytical Dose Engine for Spot Scanning Proton Delivery Systems [published online ahead of print 2019/11/23]. *Technol Cancer Res Treat.* 2019;18:1533033819887182.
71. Deng W, Ding X, Younkin JE, et al. Hybrid 3D analytical linear energy transfer calculation algorithm based on precalculated data from Monte Carlo simulations [published online ahead of print 2019/11/24]. *Med Phys.* 2020;47(2):745-752.
72. Holmes J, Shen J, Patel SH, et al. Collimating individual beamlets in pencil beam scanning proton therapy, a dosimetric investigation. *Front Oncol.* 2022;12.
73. Moignier A, Gelover E, Wang D, et al. Improving Head and Neck Cancer Treatments Using Dynamic Collimation in Spot Scanning Proton Therapy [published online ahead of print 2016/03/01]. *Int J Part Ther.* 2016;2(4):544-554.
74. Hyer DE, Hill PM, Wang D, Smith BR, Flynn RT. A dynamic collimation system for penumbra reduction in spot-scanning proton therapy: proof of concept [published online ahead of print 2014/09/05]. *Med Phys.* 2014;41(9):091701.
75. Nelson NP, Culberson WS, Hyer DE, et al. Development and validation of the Dynamic Collimation Monte Carlo simulation package for pencil beam scanning proton therapy. *Medical Physics.* 2021;48(6):3172-3185.
76. Kueng R, Guyer G, Volken W, et al. Development of an extended Macro Monte Carlo method for efficient and accurate dose calculation in magnetic fields. *Medical Physics.* 2020;47(12):6519-6530.
77. Agostinelli S, Allison J, Amako K, et al. Geant4—a simulation toolkit. *Nuclear Instruments and Methods in Physics Research Section A: Accelerators, Spectrometers, Detectors and Associated Equipment.* 2003;506(3):250-303.
78. Battistoni G, Bauer J, Boehlen TT, et al. The FLUKA Code: An Accurate Simulation Tool for Particle Therapy. *Frontiers in Oncology.* 2016;6.
79. Waters LS. MCNPX user's manual. *Los Alamos National Laboratory.* 2002;124.
80. Fippel M, Soukup M. A Monte Carlo dose calculation algorithm for proton therapy. *Medical Physics.* 2004;31(8):2263-2273.
81. Tseung HWC, Ma J, Beltran C. A fast GPU-based Monte Carlo simulation of proton transport with detailed modeling of nonelastic interactions. *Medical Physics.* 2015;42(6):2967-2978.

82. Souris K, Lee JA, Sterpin E. Fast multipurpose Monte Carlo simulation for proton therapy using multi- and many-core CPU architectures. *Medical Physics.* 2016;43(4):1700-1712.
83. Jia X, Schümann J, Paganetti H, Jiang SB. GPU-based fast Monte Carlo dose calculation for proton therapy. *Physics in Medicine & Biology.* 2012;57(23):7783.
84. Schiavi A, Senzacqua M, Pioli S, et al. Fred: a GPU-accelerated fast-Monte Carlo code for rapid treatment plan recalculation in ion beam therapy. *Physics in Medicine & Biology.* 2017;62(18):7482.
85. Yepes P, Randeniya S, Taddei PJ, Newhauser WD. Monte Carlo fast dose calculator for proton radiotherapy: application to a voxelized geometry representing a patient with prostate cancer. *Physics in Medicine & Biology.* 2009;54(1):N21.
86. Fix MK, Frei D, Volken W, Born EJ, Aebersold DM, Manser P. Macro Monte Carlo for dose calculation of proton beams. *Physics in Medicine & Biology.* 2013;58(7):2027.
87. Shan J, Feng H, Morales DH, et al. Virtual particle Monte Carlo: A new concept to avoid simulating secondary particles in proton therapy dose calculation. *Medical Physics.* 2022;49(10):6666-6683.
88. Lee H, Shin J, Verburg JM, et al. MOQUI: an open-source GPU-based Monte Carlo code for proton dose calculation with efficient data structure. *Physics in Medicine & Biology.* 2022;67(17):174001.
89. Li JS, Shahine B, Fourkal E, Ma CM. A particle track-repeating algorithm for proton beam dose calculation. *Physics in Medicine & Biology.* 2005;50(5):1001.
90. Deng W, Younkin JE, Souris K, et al. Technical Note: Integrating an open source Monte Carlo code "MCsquare" for clinical use in intensity-modulated proton therapy [published online ahead of print 2020/03/11]. *Med Phys.* 2020;47(6):2558-2574.
91. Mein S, Choi K, Kopp B, et al. Fast robust dose calculation on GPU for high-precision 1H, 4He, 12C and 16O ion therapy: the FRoG platform. *Scientific Reports.* 2018;8(1):14829.
92. Verhaegen F, Palmans H. A systematic Monte Carlo study of secondary electron fluence perturbation in clinical proton beams (70–250 MeV) for cylindrical and spherical ion chambers. *Medical Physics.* 2001;28(10):2088-2095.
93. Berger MJ. Monte Carlo calculation of the penetration and diffusion of fast charged particles. *Methods in Computational Physics.* 1963;135.
94. Kawrakow I. Accurate condensed history Monte Carlo simulation of electron transport. I. EGSnrc, the new EGS4 version. *Medical physics.* 2000;27(3):485-498.
95. Salvat F, Fernández-Varea JM, Sempau J. PENELOPE-2006: A code system for Monte Carlo simulation of electron and photon transport. Paper presented at: Workshop proceedings2006.
96. Lysakovski P, Ferrari A, Tessonnier T, et al. Development and benchmarking of a Monte Carlo dose engine for proton radiation therapy. *Frontiers in Physics.* 2021.655.
97. Collaboration G. Physics reference manual. *Version: geant4.* 2020;9(0).
98. Wilfried S, Thomas B, Wolfgang S. Correlation between CT numbers and tissue parameters needed for Monte Carlo simulations of clinical dose distributions. *Physics in Medicine & Biology.* 2000;45(2):459.
99. Group PD, Zyla P, Barnett R, et al. Review of particle physics. *Progress of Theoretical and Experimental Physics.* 2020;2020(8):083C001.
100. ESTAR, PSTAR, and ASTAR: Computer Programs for Calculating Stopping-Power and Range Tables for Electrons, Protons, and Helium Ions (version 1.2.3). 2005. http://physics.nist.gov/Star. Accessed 2023.6.1.
101. Seltzer SM, Berger MJ. Energy loss straggling of protons and mesons: Tabulation of the Vavilov distribution. *Studies in penetration of charged particles in matter.* 1964. (39):187.
102. Bohr N. *The penetration of atomic particles through matter.* Munksgaard Copenhagen; 1960.
103. Vavilov P. Ionization losses of high-energy heavy particles. *Soviet Phys JETP.* 1957;5.

104. Chibani O. New algorithms for the Vavilov distribution calculation and the corresponding energy loss sampling. *IEEE Transactions on Nuclear Science.* 1998;45(5):2288-2292.
105. Moliere G. Theorie der Streuung schneller geladener Teilchen II Mehrfach-und Vielfachstreuung. *Zeitschrift für Naturforschung A.* 1948;3(2):78-97.
106. Rossi B, Greisen K. Cosmic-Ray Theory. *Reviews of Modern Physics.* 1941;13(4):240-309.
107. Highland VL. Some practical remarks on multiple scattering. *Nuclear Instruments and Methods.* 1975;129(2):497-499.
108. Kuhn SE, Dodge GE. A fast algorithm for Monte Carlo simulations of multiple Coulomb scattering. *Nuclear Instruments and Methods in Physics Research Section A: Accelerators, Spectrometers, Detectors and Associated Equipment.* 1992;322(1):88-92.
109. Arndt RA, Strakovsky II, Workman RL. Nucleon-nucleon elastic scattering to 3 GeV. *Physical Review C.* 2000;62(3):034005.
110. Cugnon J, L'Hôte D, Vandermeulen J. Simple parametrization of cross-sections for nuclear transport studies up to the GeV range. *Nuclear Instruments and Methods in Physics Research Section B: Beam Interactions with Materials and Atoms.* 1996;111(3):215-220.
111. Trkov A, Brown DA. *ENDF-6 Formats Manual: Data Formats and Procedures for the Evaluated Nuclear Data Files.* United States2018.
112. Tripathi RK, Wilson JW, Cucinotta FA. A method for calculating proton–nucleus elastic cross-sections. *Nuclear Instruments and Methods in Physics Research Section B: Beam Interactions with Materials and Atoms.* 2002;194(3):229-236.
113. Ranft J. ESTIMATION OF RADIATION PROBLEMS AROUND HIGH ENERGY ACCELERATORS USING CALCULATIONS OF THE HADRONIC CASCADE IN MATTER. *Particle Accel 3: No 3, 129-61 (Jun 1972).* 1972.Medium: X 2009-2012-2014.
114. Paganetti H. Nuclear interactions in proton therapy: dose and relative biological effect distributions originating from primary and secondary particles. *Physics in Medicine & Biology.* 2002;47(5):747.
115. Sterpin E, Sorriaux J, Vynckier S. Extension of PENELOPE to protons: Simulation of nuclear reactions and benchmark with Geant4. *Medical Physics.* 2013;40(11):111705.
116. Verburg JM, Seco J. Proton range verification through prompt gamma-ray spectroscopy. *Physics in Medicine & Biology.* 2014;59(23):7089.
117. Wrońska A, for the SiFi CCg. Prompt gamma imaging in proton therapy - status, challenges and developments. *Journal of Physics: Conference Series.* 2020;1561(1):012021.
118. Verburg JM, Shih HA, Seco J. Simulation of prompt gamma-ray emission during proton radiotherapy. *Physics in Medicine & Biology.* 2012;57(17):5459.
119. Liu R, Charyyev S, Wahl N, et al. An Integrated Physical Optimization Framework for Proton Stereotactic Body Radiation Therapy FLASH Treatment Planning Allows Dose, Dose Rate, and Linear Energy Transfer Optimization Using Patient-Specific Ridge Filters. *International Journal of Radiation Oncology*Biology*Physics.* 2023;116(4):949-959.
120. Liu W, Feng H, Taylor PA, et al. Proton Pencil-Beam Scanning Stereotactic Body Radiation Therapy and Hypofractionated Radiation Therapy for Thoracic Malignancies: Patterns of Practice Survey and Recommendations for Future Development from NRG Oncology and PTCOG. *International Journal of Radiation Oncology*Biology*Physics.* 2024. doi: https://doi.org/10.1016/j.ijrobp.2024.01.216.
121. Volken W, Frei D, Manser P, Mini R, Born EJ, Fix MK. An integral conservative gridding-algorithm using Hermitian curve interpolation. *Physics in Medicine & Biology.* 2008;53(21):6245.
122. Li Y, Zhang X, Mohan R. An efficient dose calculation strategy for intensity modulated proton therapy [published online ahead of print 2011/01/26]. *Phys Med Biol.* 2011;56(4):N71-84.

123. Quan M, Liu W, Wu R, et al. Preliminary evaluation of multi-field and single-field optimization for the treatment planning of spot-scanning proton therapy of head and neck cancer *Med Phys.* 2013;40:081709.
124. Matney J, Park PC, Bluett J, et al. Effects of respiratory motion on passively scattered proton therapy versus intensity modulated photon therapy for stage III lung cancer: are proton plans more sensitive to breathing motion? [published online ahead of print 2013/10/01]. *International journal of radiation oncology, biology, physics.* 2013;87(3):576-582.
125. Matney JE, Park PC, Li H, et al. Perturbation of water-equivalent thickness as a surrogate for respiratory motion in proton therapy [published online ahead of print 2016/04/14]. *J Appl Clin Med Phys.* 2016;17(2):368-378.
126. Younkin J, Bues M, Keole S, Stoker J, Shen J. Multiple Energy Extraction Reduces Beam Delivery Time for a Synchrotron-Based Proton Spot-Scanning System. *Medical Physics.* 2017;44(6):2872-2872.
127. Yepes PP, Mirkovic D, Taddei PJ. A GPU implementation of a track-repeating algorithm for proton radiotherapy dose calculations. *Physics in Medicine & Biology.* 2010;55(23):7107.
128. Jahnke L, Fleckenstein J, Wenz F, Hesser J. GMC: a GPU implementation of a Monte Carlo dose calculation based on Geant4. *Physics in Medicine & Biology.* 2012;57(5):1217.
129. Du X, Liu T, Ji W, Xu XG, Brown FB. Evaluation of vectorized Monte Carlo algorithms on GPUs for a neutron Eigenvalue problem. Conference: M and C 2013: 2013 International Conference on Mathematics and Computational Methods Applied to Nuclear Science and Engineering, Sun Valley, ID (United States), 5-9 May 2013; Other Information: Country of input: France; 14 refs.; Related Information: In: Proceedings of the 2013 International Conference on Mathematics and Computational Methods Applied to Nuclear Science and Engineering - M and C 2013| 3016 p.; 2013; United States.
130. Wang Q, Adair A, Deng Y, et al. A track repeating algorithm for intensity modulated carbon ion therapy. *Physics in Medicine & Biology.* 2019;64(9):095026.
131. Renaud M-A, Roberge D, Seuntjens J. Latent uncertainties of the precalculated track Monte Carlo method. *Medical Physics.* 2015;42(1):479-490.
132. Deng W, Ding X, Younkin JE, et al. Hybrid 3D analytical linear energy transfer calculation algorithm based on precalculated data from Monte Carlo simulations [published online ahead of print 2019/11/24]. *Med Phys.* 2019. doi: 10.1002/mp.13934.
133. Deng W, Yang Y, Liu C, et al. A Critical Review of LET-Based Intensity-Modulated Proton Therapy Plan Evaluation and Optimization for Head and Neck Cancer Management. *International Journal of Particle Therapy.* 2021;8(1):36-49.
134. Li Y, Tian Z, Song T, et al. A new approach to integrate GPU-based Monte Carlo simulation into inverse treatment plan optimization for proton therapy [published online ahead of print 2016/12/20]. *Phys Med Biol.* 2017;62(1):289-305.
135. Pross D, Wuyckens S, Deffet S, Sterpin E, Lee J, Souris K. Beamlet-free optimization for Monte Carlo based treatment planning in proton therapy. *arXiv preprint arXiv:230408105.* 2023.
136. Voulodimos A, Doulamis N, Doulamis A, Protopapadakis E. Deep learning for computer vision: A brief review. *Computational intelligence and neuroscience.* 2018;2018.
137. Chai J, Zeng H, Li A, Ngai EW. Deep learning in computer vision: A critical review of emerging techniques and application scenarios. *Machine Learning with Applications.* 2021;6:100134.
138. O'Mahony N, Campbell S, Carvalho A, et al. Deep learning vs. traditional computer vision. Paper presented at: Advances in Computer Vision: Proceedings of the 2019 Computer Vision Conference (CVC), Volume 1 12020.
139. Sahiner B, Pezeshk A, Hadjiiski LM, et al. Deep learning in medical imaging and radiation therapy. *Medical physics.* 2019;46(1):e1-e36.

140. Neishabouri A, Wahl N, Mairani A, Kothe U, Bangert M. Long short-term memory networks for proton dose calculation in highly heterogeneous tissues [published online ahead of print 2020/12/18]. *Med Phys.* 2021;48(4):1893-1908.
141. Zhang X, Hu Z, Zhang G, Zhuang Y, Wang Y, Peng H. Dose calculation in proton therapy using a discovery cross-domain generative adversarial network (DiscoGAN) [published online ahead of print 2021/02/18]. *Med Phys.* 2021;48(5):2646-2660.
142. Pastor-Serrano O, Perko Z. Millisecond speed deep learning based proton dose calculation with Monte Carlo accuracy [published online ahead of print 2022/04/22]. *Phys Med Biol.* 2022;67(10).
143. Wu C, Nguyen D, Xing Y, et al. Improving Proton Dose Calculation Accuracy by Using Deep Learning [published online ahead of print 2021/03/01]. *Mach Learn Sci Technol.* 2021;2(1).
144. Bai T, Wang B, Nguyen D, Jiang S. Deep dose plugin: towards real-time Monte Carlo dose calculation through a deep learning-based denoising algorithm. *Machine Learning: Science and Technology.* 2021;2(2):025033.
145. Neph R, Lyu Q, Huang Y, Yang YM, Sheng K. DeepMC: a deep learning method for efficient Monte Carlo beamlet dose calculation by predictive denoising in magnetic resonance-guided radiotherapy. *Physics in Medicine & Biology.* 2021;66(3):035022.
146. Zhang G, Chen X, Dai J, Men K. A plan verification platform for online adaptive proton therapy using deep learning-based Monte–Carlo denoising. *Physica Medica.* 2022;103:18-25.
147. Javaid U, Souris K, Huang S, Lee JA. Denoising proton therapy Monte Carlo dose distributions in multiple tumor sites: A comparative neural networks architecture study. *Physica Medica.* 2021;89:93-103.
148. Zhang X, Zhang H, Wang J, et al. Deep learning‐based fast denoising of Monte Carlo dose calculation in carbon ion radiotherapy. *Medical Physics.* 2023;50(12):7314-7323.
149. Peng Z, Shan H, Liu T, Pei X, Wang G, Xu XG. MCDNet–a denoising convolutional neural network to accelerate Monte Carlo radiation transport simulations: A proof of principle with patient dose from x-ray CT imaging. *IEEE Access.* 2019;7:76680-76689.
150. Yan D, Vicini F, Wong J, Martinez A. Adaptive radiation therapy. *Physics in Medicine & Biology.* 1997;42(1):123.
151. Lim-Reinders S, Keller BM, Al-Ward S, Sahgal A, Kim A. Online adaptive radiation therapy. *International Journal of Radiation Oncology* Biology* Physics.* 2017;99(4):994-1003.
152. Nguyen D, Jia X, Sher D, et al. 3D radiotherapy dose prediction on head and neck cancer patients with a hierarchically densely connected U-net deep learning architecture [published online ahead of print 2019/02/01]. *Phys Med Biol.* 2019;64(6):065020.
153. Nguyen D, Long T, Jia X, et al. A feasibility study for predicting optimal radiation therapy dose distributions of prostate cancer patients from patient anatomy using deep learning [published online ahead of print 2019/02/02]. *Sci Rep.* 2019;9(1):1076.
154. Fan J, Wang J, Chen Z, Hu C, Zhang Z, Hu W. Automatic treatment planning based on three-dimensional dose distribution predicted from deep learning technique [published online ahead of print 2018/11/02]. *Med Phys.* 2019;46(1):370-381.
155. Zhang L, Holmes JM, Liu Z, et al. Beam mask and sliding window-facilitated deep learning-based accurate and efficient dose prediction for pencil beam scanning proton therapy [published online ahead of print 20230925]. *Med Phys.* 2023. doi: 10.1002/mp.16758.
156. Zhang L, Holmes JM, Liu Z, et al. Noisy probing dose facilitated dose prediction for pencil beam scanning proton therapy: physics enhances generalizability. *arXiv preprint arXiv:231200975.* 2023.
157. Holmes J, Shen J, Shan J, et al. Technical note: Evaluation and second check of a commercial Monte Carlo dose engine for small-field apertures in pencil beam scanning proton therapy. *Medical Physics.* 2022;49(5):3497-3506.

158. Feng H, Holmes J, Vora SA, et al. Modelling small block aperture in an in-house developed GPU-accelerated Monte Carlo-based dose engine for pencil beam scanning proton therapy. *Physics in Medicine & Biology.* 2023.
159. Dowdell SJ, Clasie B, Depauw N, et al. Monte Carlo study of the potential reduction in out-of-field dose using a patient-specific aperture in pencil beam scanning proton therapy. *Physics in Medicine & Biology.* 2012;57(10):2829.
160. Rana S, Storey M, Manthala Padannayil N, et al. Investigating the utilization of beam-specific apertures for the intensity-modulated proton therapy (IMPT) head and neck cancer plans. *Medical Dosimetry.* 2021;46(2):e7-e11.
161. Hyer DE, Bennett LC, Geoghegan TJ, Bues M, Smith BR. Innovations and the Use of Collimators in the Delivery of Pencil Beam Scanning Proton Therapy. *International Journal of Particle Therapy.* 2021;8(1):73-83.
162. Ciocca M, Magro G, Mastella E, et al. Design and commissioning of the non-dedicated scanning proton beamline for ocular treatment at the synchrotron-based CNAO facility [published online ahead of print 2019/01/20]. *Med Phys.* 2019;46(4):1852-1862.
163. Fellin F, Righetto R, Braccelli M, et al. Is It Beneficial to Use Apertures in Proton Radiosurgery with a Scanning Beam? A Dosimetric Comparison. *International Journal of Radiation Oncology, Biology, Physics.* 2019;105(1):E760-E761.
164. Maes D, Regmi R, Taddei P, et al. Parametric characterization of penumbra reduction for aperture-collimated pencil beam scanning (PBS) proton therapy. *Biomedical Physics & Engineering Express.* 2019;5(3):035002.
165. Bäumer C, Fuentes C, Janson M, Matic A, Timmermann B, Wulff J. Stereotactical fields applied in proton spot scanning mode with range shifter and collimating aperture. *Physics in Medicine & Biology.* 2019;64(15):155003.
166. Paganetti H, Jiang H, Parodi K, Slopsema R, Engelsman M. Clinical implementation of full Monte Carlo dose calculation in proton beam therapy. *Physics in Medicine & Biology.* 2008;53(17):4825.
167. Parodi K, Ferrari A, Sommerer F, Paganetti H. Clinical CT-based calculations of dose and positron emitter distributions in proton therapy using the FLUKA Monte Carlo code. *Physics in Medicine & Biology.* 2007;52(12):3369.
168. Schneider U, Agosteo S, Pedroni E, Besserer J. Secondary neutron dose during proton therapy using spot scanning. *International Journal of Radiation Oncology* Biology* Physics.* 2002;53(1):244-251.
169. Aso T, Kimura A, Tanaka S, et al. Verification of the dose distributions with GEANT4 simulation for proton therapy. *IEEE transactions on Nuclear Science.* 2005;52(4):896-901.
170. Paganetti H. Monte Carlo calculations for absolute dosimetry to determine machine outputs for proton therapy fields. *Physics in Medicine & Biology.* 2006;51(11):2801.
171. Rogers DWO. Fifty years of Monte Carlo simulations for medical physics*. *Physics in Medicine & Biology.* 2006;51(13):R287.
172. Okada S, Murakami K, Incerti S, Amako K, Sasaki T. MPEXS-DNA, a new GPU-based Monte Carlo simulator for track structures and radiation chemistry at subcellular scale. *Medical Physics.* 2019;46(3):1483-1500.
173. Keyes RW, Romano C, Arnold D, Luan S. Radiation therapy calculations using an on-demand virtual cluster via cloud computing. *arXiv preprint arXiv:10095282.* 2010.
174. Green A, Aitkenhead A, Owen HL, Mackay RI. Fast Monte Carlo proton treatment plan validation in the Google Cloud. *Physics in Medicine and Biology.* 2015.
175. Wang J-L, Wu X-G, Li Z-F, et al. Prompt gamma spectroscopy retrieval algorithm for element and density measurements accelerated by cloud computing. *Frontiers in Physics.* 2022;10:1097.

176. Fracchiolla F, Engwall E, Janson M, et al. Clinical validation of a GPU-based Monte Carlo dose engine of a commercial treatment planning system for pencil beam scanning proton therapy. *Physica Medica.* 2021;88:226-234.
177. Souris K, Lee JA, Sterpin E. Fast multipurpose Monte Carlo simulation for proton therapy using multi- and many-core CPU architectures [published online ahead of print 2016/04/03]. *Med Phys.* 2016;43(4):1700.
178. Gajewski J, Garbacz M, Chang C-W, et al. Commissioning of GPU–Accelerated Monte Carlo Code FRED for Clinical Applications in Proton Therapy. *Frontiers in Physics.* 2021;8.
179. Lysakovski P, Ferrari A, Tessonnier T, et al. Development and Benchmarking of a Monte Carlo Dose Engine for Proton Radiation Therapy. *Frontiers in Physics.* 2021;9.
180. Paganetti H, Jiang H, Adams JA, Chen GT, Rietzel E. Monte Carlo simulations with time-dependent geometries to investigate effects of organ motion with high temporal resolution [published online ahead of print 2004/10/07]. *International journal of radiation oncology, biology, physics.* 2004;60(3):942-950.
181. Rosu M, Hugo GD. Advances in 4D radiation therapy for managing respiration: part II - 4D treatment planning [published online ahead of print 2012/07/17]. *Zeitschrift fur medizinische Physik.* 2012;22(4):272-280.
182. Manser P, Frauchiger D, Frei D, Volken W, Terribilini D, Fix MK. Dose calculation of dynamic trajectory radiotherapy using Monte Carlo [published online ahead of print 2018/04/11]. *Zeitschrift fur medizinische Physik.* 2019;29(1):31-38.
183. Kang M, Ding X, Rong Y. FLASH instead of proton arc therapy is a more promising advancement for the next generation proton radiotherapy [published online ahead of print 2023/07/11]. *J Appl Clin Med Phys.* 2023;24(8):e14091.
184. Jeans MEd M, Elizabeth B., Grams PhD MP, Spraker MD P, Matthew B., Zeman PhD EM, Ma MD DJ. Grid Therapy. In: *Principles and Practice of Particle Therapy.* doi: https://doi.org/10.1002/9781119707530.ch102022:137-149.
185. Pepin MD, Tryggestad E, Wan Chan Tseung HS, Johnson JE, Herman MG, Beltran C. A Monte-Carlo-based and GPU-accelerated 4D-dose calculator for a pencil beam scanning proton therapy system [published online ahead of print 2018/09/12]. *Med Phys.* 2018;45(11):5293-5304.
186. Paganetti H. Mechanisms and Review of Clinical Evidence of Variations in Relative Biological Effectiveness in Proton Therapy [published online ahead of print 2021/08/19]. *International journal of radiation oncology, biology, physics.* 2022;112(1):222-236.